\documentclass[runningheads]{llncs}

\usepackage{graphicx}
\usepackage{amsfonts}
\usepackage{latexsym,amssymb,amsmath}

\usepackage[ruled]{algorithm}
\usepackage[noend]{algpseudocode}

\usepackage{comment}
\usepackage{epstopdf}
\usepackage[hidelinks]{hyperref}
\usepackage{microtype}

\usepackage[appendix=inline]{apxproof}

\newtheoremrep{lemma}{Lemma}[section]{\bfseries}{\itshape}
\newtheoremrep{theorem}{Theorem}[section]{\bfseries}{\itshape}

\begin{document}
\title{
Delivery to Safety with Two Cooperating Robots\thanks{Research supported in part by NSERC Discovery grant}
}

\author{
Jared Coleman\inst{2} 
\and
Evangelos Kranakis\inst{3} 
\and
Danny Krizanc\inst{4}
\and
Oscar Morales-Ponce\inst{1} 
}

\authorrunning{J. Coleman et al.}

\institute{
    Department of Computer Engineering and Computer Science, California State University, Long Beach, Long Beach, CA.
    \email{Oscar.MoralesPonce@csulb.edu}
    \and
    Department of Computer Science, University of Southern California, Los Angeles, CA, USA.
    \email{jaredcol@usc.edu}
    \and
    School of Computer Science, Carleton University, Ottawa, Ontario, Canada.
    \email{kranakis@scs.carleton.ca}
    \and
    Department of Mathematics \& Computer Science, Wesleyan University, Middletown CT, USA.
    \email{dkrizanc@wesleyan.edu}
}

\maketitle

\begin{abstract}
Two cooperating, autonomous mobile robots with arbitrary nonzero max speeds are placed at arbitrary initial positions in the plane. A remotely detonated bomb is discovered at some source location and must be moved to a safe distance away from its initial location as quickly as possible. In the {\em Bomb Squad} problem, the robots cooperate by communicating face-to-face in order to pick up the bomb from the source and carry it away to the boundary of a disk centered at the source in the shortest possible time. The goal is to specify trajectories which define the robots' paths from start to finish and their meeting points which enable face-to-face collaboration by exchanging information and passing the bomb from robot to robot.

We design algorithms reflecting the robots' knowledge about orientation and each other's speed and location. In the offline case, we design an optimal algorithm. For the limited knowledge cases, we provide online algorithms which consider robots' level of agreement on orientation as per {\tt OneAxis} and {\tt NoAxis} models, and knowledge of the boundary as per {\tt Visible, Discoverable,} and {\tt Invisible}. In all cases, we provide upper and lower bounds for the competitive ratios of the online problems.
\end{abstract}

\section{Introduction}\label{sec:intro}
A remotely detonated bomb is located at the center of some critical zone. 
Since the time of detonation is unknown, the bomb must be removed as quickly as possible from the critical zone by two autonomous mobile robots. 
How can these robots, each with their own speeds and initial location, collaborate to carry the bomb out of the critical zone as quickly as possible?
We assume the bomb is initially located at a point $S$ (the source) and must be transported at least distance $D$ (called the {\em critical distance}) away from the source. The critical 
distance defines a disk centered at $S$ of radius $D$. Each robot has its own maximum speed and the bomb can be passed from robot to robot in a face-to-face communication exchange. We refer to this as the {\em Bomb Squad} problem.
The perimeter of the disk centered at $S$ and radius $D$ is also called the {\em boundary} and encloses the critical zone which must be rid of the bomb.

In the sequel, we study various versions of the Bomb Squad problem which depend on what knowledge the collaborating robots have regarding the location of the other robot and the boundary. We are interested in designing both offline (full knowledge) and online (limited knowledge) algorithms that describe the trajectories and collaboration of the participating robots.

\subsection{Model, Notation, and Preliminaries}

There are two autonomous mobile robots, $r_1$ and $r_2$, with maximum speeds $v_1$ and $v_2$ initially placed in the plane distances $d_1$ and $d_2$ from the source, respectively. 
We use the standard mobility model for the robots. 
At any time, they may stop and start, change direction/speed, and carry the bomb when they decide to do so. A robot trajectory is a continuous function $f: [0, T] \to \mathbb{R}^2$ such that $f(t)$ is the location of the robot at time $t$ and $T$ is the duration of a robot's trajectory. If the robot's speed cannot exceed $v$ then $\| f(t) - f(t^\prime) \|_2 \leq v |t-t'|$, for all $0 \leq t,t^\prime \leq T$, where $\| \cdot \|_2$ denotes the Euclidean norm in the plane $\mathbb{R}^2$. 

Robots may collect information as they traverse their trajectories. 
Moreover, they may exchange information only when they are collocated (also known as F2F model). 
When collocated, they may compare their speeds and decide which robot is faster.
They can recognize the bomb initially placed at location $S$ and can carry it around and pass it from robot to robot without their speed being affected.

We assume robots have a common unit of distance.
We consider both the offline and online settings. In the offline setting, all information regarding the robots (their initial positions and speeds) is available and an algorithm provides robot trajectories and a sequence of robot meetings that relay the bomb from the source to the boundary in optimal time.  
 
In the online setting, a robot has limited knowledge of the other robot's location and the critical distance. We consider both {\tt OneAxis} and {\tt NoAxis} 
(or {\tt Disoriented}) models  (see~\cite{flocchini2019distributed}). In the {\tt OneAxis} model  
robots agree on a single axis and direction (i.e. North). In the {\tt NoAxis} model, we say robots are {\em disoriented} and do not agree on any axis or direction.
With respect to knowledge of the critical distance, we consider three models:
\begin{enumerate}
    \item {\tt VisibleBoundary}: the boundary is always visible and thus the critical distance $D$ is known by all robots.
    \item {\tt DiscoverableBoundary}: the boundary (and thus the critical distance) is not known ahead of time but is ``discoverable''. Robots can {\em discover} the boundary (and the critical distance) by visiting any point on the boundary or by encountering another robot which has already discovered it.
    \item {\tt InvisibleBoundary}: the boundary is completely invisible and robots have no knowledge of whether or not they've already visited a point on the boundary.
\end{enumerate} 
Each of these models has intuitive inspiration from the bomb-squad scenario.
The {\tt VisibleBoundary} model considers the situation where a safe distance is known ahead of time, while the {\tt DiscoverableBoundary} model considers a situation where a boundary --- physical (i.e. a fence or a wall) or abstract (i.e. a border, patrol line, maximum communication distance) --- must be discovered by the robots.
Finally, the {\tt InvisibleBoundary} model considers the situation where a safe distance is not known by the robots (i.e. they don't know the detonation radius of the bomb).
In this case, the goal is to deliver the bomb to an unknown radius as quickly as possible.
Interestingly, each of these models yields unique algorithms with different competitive ratios.

In all of our algorithms, both robots start at the same time from arbitrary locations in the plane. 
The delivery time $T_A(I)$ of an algorithm $A$ solving the Bomb Squad problem is the time it takes the algorithm $A$ to deliver the bomb to the boundary for an instance $I$ of the problem (a source location $S$, critical distance $D$, and robots' initial positions and maximum speeds). 
If $T_{opt}(I)$ is the optimal time of an offline algorithm for the same instance $I$, then the competitive ratio of an online algorithm $A$ is defined by the ratio $CR_A := \sup_{I} T_A(I)/T_{opt}(I)$. 
If ${\cal A} $ is a class of algorithms solving an online version of the Bomb Squad problem, then its competitive ratio is defined by $CR_{{\cal A}}:= \inf_{A \in {\cal A}} CR_A$. 
Usually, the subscripts will be omitted since the online version of the problem will be easily understood from the context.


In proving upper bounds on the competitive ratio, if the faster robot cannot arrive at $S$ before the slow robot then we may restrict our attention to the case where the slow robot starts at the source. 
We state this useful claim as a lemma.
\begin{lemmarep}\label{lm:start}
    Consider any online algorithm solving an online version of the Bomb Squad problem. Assume that the faster robot cannot arrive at $S$ before the slow robot does. If $c$ is an upper bound on the competitive ratio of the algorithm for all instances in which the slow robot starts at the source, then $c$ is also an upper bound on the competitive ratio for that algorithm. 
\end{lemmarep}
\begin{proof}
    Consider any online algorithm $A$ solving the given optimization problem. 
    Recall that the robots can communicate only F2F. 
    Let $t$ be the time it takes the slow robot to move from its starting position to the source $S$. 
    The hypothesis of the lemma implies after time $t$ has passed the faster robot cannot arrive at the source. 
    Therefore algorithm $A$ is split into two parts: the first part takes time $t$ and the second part is an algorithm $A_t$ that assumes that the slow robot started at $S$. Now observe that 
    $$
    CR_A = \frac{T_A}{T_{Opt}} = \frac{t+T_{A_t}}{t+T_{Opt_t}} \leq \frac{T_{A_t}}{T_{Opt_t}} \leq c ,
    $$
    where $Opt_t$ is the optimal algorithm after time $t$ has passed and the slow robot starts at $S$.
    ~\qed
\end{proof}

\subsection{Related work}\label{sec:related}
The {\em Bomb Squad} problem is closely related to the message delivery problem with a set of robots.
In that problem, the source and destination are predefined and robots jointly work to deliver the message. 
Two different objective functions have been studied. 
The first assumes that the robots have limited battery and consequently the objective function is to minimize the maximum movement (minmax).
The second is to minimize the time to deliver the message.
Anaya et al.~\cite{chalopin2013data} study a more general minmax problem where the message must be delivered to many destinations. 
The authors show that the decision problem is NP-hard and provide a 2 approximation algorithm.

Chalopin et al.~\cite{chalopin2014data} study the minmax problem on a line and show that the decision problem is NP-Complete for instances where all input values are integers. The authors also provide an algorithm for delivering the message that runs in $O(d^2 \cdot n^{1+4 \log d})$ time where $d$ is the distance between the source and destination for the general case. Coleman et al.~\cite{coleman2021pony} study the broadcast and unicast versions of the problem on a line and present optimal offline algorithms and online algorithms with optimal constant competitive ratio. In~\cite{czyzowicz2016communication} Czyzowicz et al. study the problem in a weighted graph and show that the problem is NP-Complete. They also show that by allowing robots to exchange energy, the problem can be solved in polynomial time.
Carvalho et al.~\cite{carvalho2021fast} also study the problem in weighted graphs. 
They provide an offline algorithm that runs in $O(k n \log n + km)$ time where $k$ is the number of robots, $n$ is the number of nodes, and $m$ the number of edges.

More recently, Coleman et al.~\cite{pony_plane} studied the point-to-point delivery problem on the plane and gave an optimal offline algorithm for two robots as well as approximation offline algorithms and online algorithms with constant competitive ratio. The delivery problem differs significantly from the problem studied in our current paper, where the goal is to reach any point on a given boundary (namely the perimeter of a disk centered at the source) as opposed to a specific destination.

The delivery problem studied in our paper focuses on the knowledge the robots have about each other as well as the environment. To this end we design algorithms for the {\tt OneAxis} and {\tt NoAxis} models. In particular, in the latter model and based on the knowledge the robots have in Subsection~\ref{sec:online_noaxis_invisibleboundary} one has to design a search algorithm that makes the robots perform a ``zigzag'' procedure in order to collect appropriate information and pass the bomb to the faster robot, if feasible, that will eventually deliver the bomb to the boundary. This has similarities to the well-known linear search algorithms proposed by Baeza-Yates~et al.~\cite{baezayates1993searching}, Beck~\cite{beck1964linear} and Bellman~\cite{bellman1963optimal}, Ahlswede et al.~\cite{ahlswede1987search}, as well as Alpern et al.~\cite{ahlswede1987search,alpern2003theory}. However, search in the previously given research works is based only on one robot while in our case we have two collaborating robots with incomplete information about the environment.

\subsection{Outline and results of the paper}
In this paper, we design and analyze algorithms for the Bomb Squad problem with two cooperating robots. In the offline case, we design an optimal algorithm that assumes robots have knowledge of their own and each other's location but does not require knowledge of each other's speed.
\begin{table}[htp]
\caption{Upper and lower bounds on the competitive ratio of online algorithms for two cooperating robots in the {\tt OneAxis} and {\tt NoAxis} models for {\tt Visible, Discoverable}, and {\tt Invisible} Boundary:}
\begin{center}
\begin{tabular}{| l | l | c | c | c |}
\hline
Axis Model & Boundary Model & Upper Bound & Lower Bound & Section\\
\hline
{\tt OneAxis} & All & $\frac{1}{7} \left( 5 + 4 \sqrt{2} \right) \approx 1.5224$ & $1.48102$ & \ref{sec:oneaxis} \\
\hline
{\tt NoAxis} & {\tt Visible} & $1+\sqrt 2$ & $1+ \sqrt 2$ & \ref{sec:online_noaxis_visibleboundary} \\
\hline
{\tt NoAxis} & {\tt Discoverable} & $\frac {15}4$ & $3$ & \ref{sec:Online Algorithm for the DiscoverableBoundary Model}\\
\hline
{\tt NoAxis} & {\tt Invisible} & $\frac{7+\sqrt{17}}2$ & $2+\sqrt 5$ & \ref{sec:online_noaxis_invisibleboundary} \\
\hline
\end{tabular}
\end{center}
\label{default}
\end{table}%
For the online case Table~\ref{default} displays upper and lower bounds on the competitive ratio for the {\tt OneAxis} and {\tt NoAxis} models, and for {\tt Visible, Discoverable}, and {\tt Invisible} Boundary as well as the specific (sub)section where the results are proved. Section~\ref{sec:Optimal Offline Algorithm} presents an optimal offline algorithm, 
Section~\ref{sec:oneaxis} presents an online algorithm for the {\tt OneAxis} model, while Section~\ref{sec:Online Algorithms for NoAxis Model} includes the results of the three Subsections for the {\tt NoAxis} model. There are many interesting open problems and in Section~\ref{sec:Conclusion} we summarize the results and discuss potential extensions and alternatives.
\section{Optimal Offline Algorithm}
\label{sec:Optimal Offline Algorithm}

Our problem may be solved optimally using Algorithm~\ref{alg:offline}.
\vspace*{-0.5cm}
\begin{algorithm}[!h]
\caption{Offline Delivery Algorithm for Two Robots}\label{alg:offline}
    \begin{algorithmic}[1]
        \State move toward $S$
        \If {arrived at $S$}
            \State pick up the bomb 
            \State move in direction of other robot
        \ElsIf {encountered other robot with bomb and other robot is slower}
            \State take the bomb from other robot
            \State move away from $S$ toward boundary
        \EndIf 
    \end{algorithmic}
\end{algorithm}
\vspace*{-0.5cm}

\begin{theorem}\label{thm:offline}
    For any two robots $r_1, r_2$ such that $v_1 \leq v_2$, the offline Algorithm~\ref{alg:offline} is optimal in that it delivers the bomb to the perimeter of the circle centered at $S$ with radius $D$ in minimum time
    \begin{align}\label{eq:p2c1}
        \min \left( 
            \frac{d_1 + D}{v_1},~
            \frac{d_2 + D}{v_2},~
            \frac{D - d_2}{v_2} + 2 \frac{d_1+d_2}{v_1+v_2}
        \right)
    \end{align}
    where $S$ is the initial location of the bomb and $d_1, d_2$ are the starting distances (from $S$) of the $r_1, r_2$, respectively.
\end{theorem}
\begin{inlineproof}
    First, observe that the cases where the fast robot can reach $S$ first or where the slow robot can deliver the bomb before the fast robot can get within a distance $D$ of $S$ are trivial and justify the first two arguments of the $\min$ term in~\eqref{eq:p2c1}.
    In each case the robot which reaches $S$ first simply completes the delivery by itself and the algorithm is optimal.
    In all other cases, the slow robot reaches $S$ first and must hand the bomb over to the fast robot at some point $M$ which then delivers it to the boundary. 
    Observe that the trajectory of the bomb itself must be a straight line since the closest point from $M$ to the perimeter of the circle must be along $SM$ (by the definition of a circle).
    
    Consider all the candidate trajectories of the bomb.
    Since it must travel a total distance of exactly $D$, the trajectory which minimizes the delivery time is clearly that which involves the faster of the two robots carrying the bomb for the greatest portion of this distance.
    In other words, if $s$ is the distance the slow robot carries the bomb before handing it over to the fast robot (Figure~\ref{fig:offline_circle_2}), then the delivery time is
    \begin{align*}
        \frac{s}{v_1} + \frac{D-s}{v_2} &= \frac{s (v_2 - v_1) + D v_1}{v_1 v_2}
    \end{align*}
    which is clearly minimized when $s$ is minimum since $v_1 \leq v_2$.
    Intuitively, this means the slow robot should carry the bomb as short a distance as possible.
    Clearly, $s$ is minimum when the slow robot moves directly toward the fast robot.
    \begin{figure}[!htb]
        \begin{center}
            \includegraphics[width=4cm]{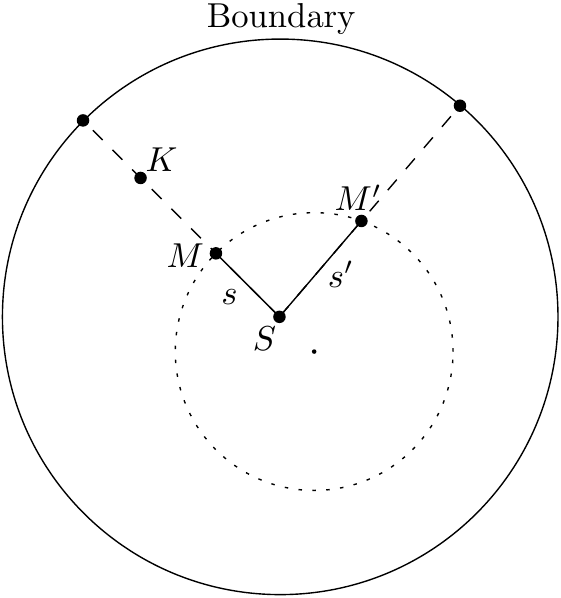}
        \end{center}
        \caption{Two candidate trajectories given by robot meeting points $M$ and $M^\prime$, where the slow robot starts at $S$ and the fast robot at $K$. Clearly, $M$ is superior since $s < s^\prime$ and the faster robot spends a larger portion of the bomb's trajectory carrying it. In other words, the bomb is moving at the faster speed $v_2$ for a larger portion of its trip to the boundary.}
        \label{fig:offline_circle_2}
    \end{figure}
    
    The delivery time for this case can then be easily written as the sum of the time for the robots to meet and the time for the fast robot to travel back to the boundary for delivery:
    \begin{align*}
        \frac{d_1+d_2}{v_1+v_2} 
            + \frac{1}{v_2}\left( D - \left( \left( \frac{d_1+d_2}{v_1+v_2} - \frac{d_1}{v_1} \right) v_1 \right) \right)
        = \frac{D - d_2}{v_2} + 2 \frac{d_1+d_2}{v_1+v_2}.
    \end{align*}
    ~\qed
\end{inlineproof}

\section{Online Algorithm for the {\tt OneAxis} Model}
\label{sec:oneaxis}

Here we assume the robots agree on a single axis and direction and can therefore choose to move along the same radius emanating from $S$.
We start by proving a lower bound.

\begin{theoremrep}\label{lm:oneaxis_lower}
    Any online algorithm for the {\tt OneAxis} model has a competitive ratio of at least $1.48102$.
\end{theoremrep}
\begin{proof} 
    Without loss of generality, let the source be at the origin $S=(0,0)$ and the critical distance $D=1$.
    Consider two robots $r_1$ and $r_2$ with speeds $v_1 = 1 / (1+x)$ and $v_2=1$ for some $x > 0$ whose value will be determined below.
    Suppose also that the slower robot $r_1$ starts at the source ($d_1=0$) and the faster robot starts at distance $x$ away from the source ($d_2=x$) in some direction to be determined below.
    Observe the movement of the bomb (which moves with speed at most $v_1$ when carried by $r_1$) during the time period $[0, x]$ and let $(x_1, y_1)$ denote the final position of the bomb after time $x$.
    Then let $r_2$'s initial position be the point at distance $x$ from $S$ in the opposite direction of $(x_1, y_1)$ (or any direction if $x_1=0$ and $y_1=0$).
    Observe that the trajectories taken by the bomb (via $r_1$) and $r_2$ cannot overlap during the time period $[0,x]$ except at $(0,0)$.
    Indeed, at time $t \leq x$, the bomb must be either at $S$, on the opposite side of $S$ as $r_2$, or on the same side and within distance $(x-t) v_1$ of $S$ (since it must be at or on the opposite side by time $x$).
    On the other hand $r_2$ is at least a distance $x - (v_2 t)$ away from $S$. 
    Multiplying out we see that $ (x-t) v_1 \leq x - (v_2 t)$ is equivalent to $1 \leq 1+x$.
    Thus, for any online algorithm, we can always place $r_2$ so that it cannot encounter the bomb before time $x$.

    First, observe the optimal offline delivery time in this case is
    \begin{align*}
        \frac{x}{1+ \frac{1}{1+x}} + \left( 1 - \left( \frac{x}{1+\frac{1}{1+x}} \cdot \frac{1}{1+x} \right) \right) = \frac{x^2+x+2}{x+2}.
    \end{align*}
    Also, observe any algorithm that does not involve both robots must take at least $1+x$ time and so the competitive ratio can be written as
    \begin{align*}
        \frac{1+x}{\frac{x^2+x+2}{x+2}}
    \end{align*}
    which has a maximum value of $\frac 17 (5+4\sqrt{2})$ at $x = \sqrt{2}$.
    Let us now consider algorithms which {\em do} involve both robots.
    Without loss of generality, suppose the bomb ends up at some position $(x_1, 0)$ (on the positive $x$-axis such that $0 \leq x_1 \leq x/(1+x)$) and thus $r_2$'s initial position is at $(-x, 0)$ (for some $x \geq 0$ yet to be determined).
    Then let $t + x$ denote the time at which $r_2$ acquires the bomb.
    Observe any algorithm for which $t > 1$ has a worse competitive ratio than the no-cooperation algorithm analyzed above.
    Then, we may assume for any optimal online algorithm $t \leq 1$ and thus the delivery time is at least $t + x$.
    Thus, one lower bound for the competitive ratio can be written
    \begin{align}
        \frac{x+t}{ \frac{x^2+x+2}{x+2} }. \label{eq:lower_1}
    \end{align}
    We can derive a second lower bound for this case though. 
    Observe the delivery time of the algorithm can be written
    \begin{align*}
        x + t + \left(1 - \sqrt{x_2^2 + y_2^2}\right)
    \end{align*}
    where $(x_2, y_2)$ is the location at which $r_2$ acquires the bomb.
    Observe, then that $-t/(1+x) \leq x_2 \leq t$ since $r_1$ cannot move the bomb further than $t/(1+x)$ from $x_1$ and $r_2$ cannot make it further than $t + x$ from its starting position in time $t$ 
    (see Figure~\ref{fig:oneaxis_lower}).
    Thus, $|x_2| \leq t$.
    \begin{figure}[!htb]
        \centering
        \includegraphics[width=0.7\textwidth]{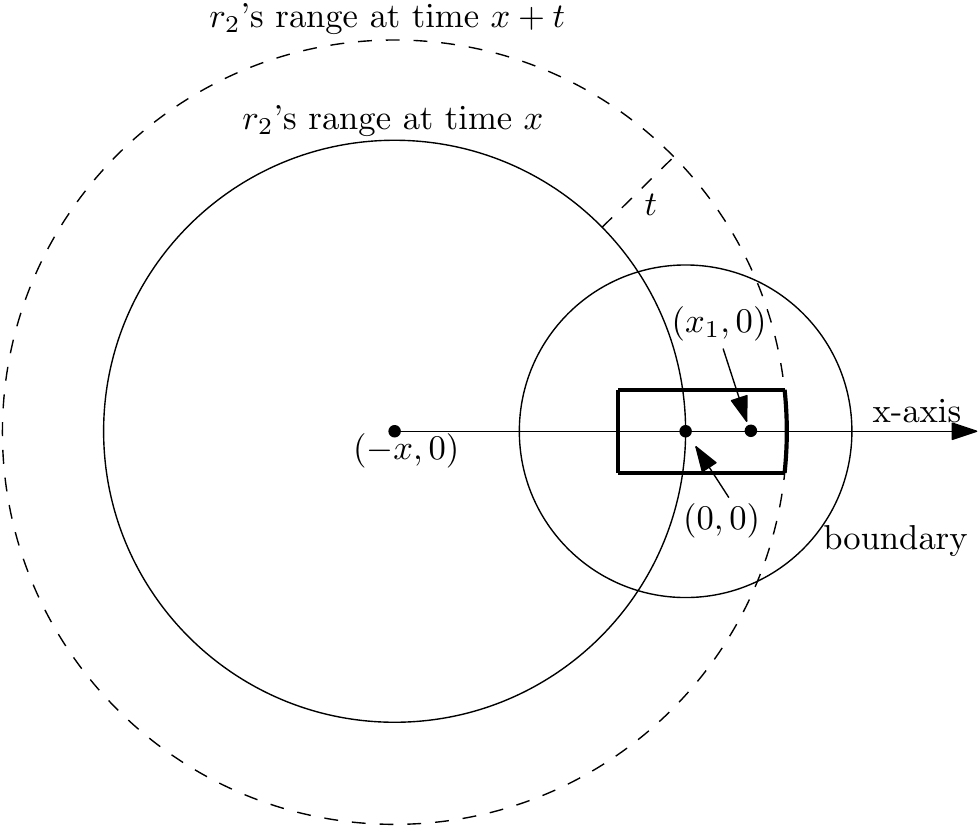}
        \caption[{\tt OneAxis} Lower Bound]{If the robots are to meet, it must be within the boundary outlined by the thick black lines, due to the positions/speeds of robots $r_1$ and $r_2$ at time $x$.}\label{fig:oneaxis_lower}
    \end{figure}
    Also observe that $|y_2| \leq t/(1+x)$ since the bomb must be on the x-axis at time $x$.
    Thus, the delivery time is at least
    \begin{align*}
        x + t + 1 - \sqrt{t^2 + \frac{t^2}{(1+x)^2}} 
        &= 1 + x + t \left( 1 - \sqrt{1 + \frac{1}{(1+x)^2}} \right)
    \end{align*}
    Thus, the competitive ratio is at least
    \begin{align}
        \frac{
            1 + x + t \left( 1 - \sqrt{1 + \frac{1}{(1+x)^2}} \right)
        }{
            \frac{x^2+x+2}{x+2}
        }. \label{eq:lower_2}
    \end{align}
    Since Lower Bound~\eqref{eq:lower_1} is increasing with respect to $t$, and Lower Bound~\eqref{eq:lower_2} is decreasing with respect to $t$ and Bounds~\eqref{eq:lower_1} and~\eqref{eq:lower_2} intersect at $t = 1/\sqrt{1+1/(1+x)^2}$, the competitive ratio is at least
    \begin{align*}
        \frac{x + \frac{1}{\sqrt{1+ \frac{1}{(1+x)^2}}}}{\frac{x^2+x+2}{x+2}}.
    \end{align*}
    Finally, we achieve the lower bound of $1.48102$ by observing the above expression is maximized at $x=1.67696$.
    ~\qed
\end{proof}

Now, we propose the following Algorithm~\ref{alg:oneaxis} and prove its competitive ratio is at most $\frac{1}{7} \left( 5+4\sqrt{2} \right)$.
\vspace*{-0.5cm}
\begin{algorithm}[!h]
    \caption{Online Delivery Algorithm for the {\tt OneAxis} Model}\label{alg:oneaxis}
    \begin{algorithmic}[1]
        \State {move toward $S$, taking the bomb from any slower robot encountered}
        \State {upon reaching $S$, move along common axis/direction away from $S$, taking the bomb from any slower robot encountered}
    \end{algorithmic}
\end{algorithm}
\vspace*{-0.5cm}

\begin{theorem}\label{lm:oneaxis_upper}
Algorithm~\ref{alg:oneaxis} has competitive ratio $\frac{1}{7} \left( 5 + 4 \sqrt{2} \right)$.
\end{theorem}
\begin{inlineproof}
    If the fast robot arrives at the center first, then clearly the algorithm is optimal (it completes the delivery entirely by itself).
    Similarly, if the optimal algorithm involves only the slow robot (i.e. the fast robot is too far away to help), the algorithm is also optimal.
    Thus, we may consider only the case where the slow robot arrives first and where an optimal offline algorithm involves cooperation between the two robots.

    Unlike in the optimal algorithm, the slow robot will not move directly toward the fast robot, since it doesn't know where it is.
    Rather, the slow robot will move along the shared axis in a previously agreed-upon direction (i.e. North).
    The fast robot will continue to move toward the source and, seeing the bomb is no longer there, begin to move along the shared axis.
    If the fast robot is fast enough, it will catch the slow robot, take the bomb, and complete the delivery.
    Otherwise, the slow robot will deliver the bomb.

    Let $d_1$ and $d_2$ be the initial distances of the slow and fast robots to the source, respectively. 
    Without loss of generality, suppose $D=1$ and the fast robot moves at speed $v_2=1$.
    By Lemma~\ref{lm:start}, setting $d_1 = 0$ cannot decrease the competitive ratio, and so a bound on the competitive ratio can be written as
    \begin{align*}
        \frac{
            \min\left\{ \frac{1}{v_1}, d_2+1 \right\}
        }{
            \frac{d_2}{v_1+1} + \left( 1 - \frac{d_2}{v_1+1} v_1 \right)
        } 
        = \frac{
            \min\left\{ \frac{1}{v_1}, d_2+1 \right\}
        }{
            d_2 \frac{1-v_1}{1+v_1} + 1
        }.
    \end{align*}
    For the first case, where $\frac{1}{v_1} \leq d_2 + 1$, we can write an upper bound by substituting $1/(d_2+1)$ for $v_1$ since $1/v_1$ decreases w.r.t $v_1$ and $\frac{1-v_1}{1+v_1}$ increases w.r.t $v_1$:
    \begin{align*}
        \frac{
            \frac{1}{v_1}
        }{
            d_2 \frac{1-v_1}{1+v_1} + 1
        } \leq 1 + \frac{2 d_2}{2+d_2+d_2^2}
    \end{align*}
    which is maximized when $d_2 = \sqrt{2}$ (giving a value of $\frac{1}{7}\left( 5+4\sqrt{2} \right)$).
    For the second case, when $\frac{1}{v_1} > d_2 + 1$, observe:
    \begin{align*}
        \frac{
            d_2+1
        }{
            \frac{d_2}{v_1+1} (1 - v_1) + 1
        } = \frac{(1+v_1) + d_2 (1+v_1)}{(1+v_1) + d_2 (1-v_1)}
        \leq \frac{1+v_1}{1+v_1(2 v_1 - 1)}
    \end{align*}
    which is maximized when $v_1 = \sqrt{2} - 1$ (giving a value of $\frac{1}{7}\left( 5+4\sqrt{2} \right)$).
    ~\qed
\end{inlineproof}

\begin{remark}
    This algorithm makes no use of the critical distance and thus applies to all three boundary-knowledge models ({\tt VisibleBoundary}, {\tt DiscoverableBoundary}, and {\tt InvisibleBoundary}).
\end{remark}

\section{Online Algorithms for the {\tt NoAxis} Model}
\label{sec:Online Algorithms for NoAxis Model}

The previous algorithms made use of a common axis and direction between the two robots.
Now we consider an even weaker model where robots are disoriented (they have no common axis or sense of direction).
We consider the three boundary-knowledge models introduced in Section~\ref{sec:intro}.

We begin with the following lemma which will be useful for the analysis of online algorithms.

\begin{lemmarep}\label{lm:meet_at_source}
Assume at the start the slow robot is at $S$. 
Any online algorithm that involves the robots meeting at any point other than $S$ cannot have constant competitive ratio.
\end{lemmarep}

\begin{proof}
    Consider any algorithm $A$. 
    Then there must be instances for which the robots must meet at some point before the final delivery of the bomb to the perimeter. 
    Indeed, if the robots do not meet, $A$ cannot have a constant competitive ratio. 

    Let the slow robot at $S$ be $r_1$ with speed $v_1$, while the other robot, say $r_2$ with speed $v_2 > v_1$, is at some other point $P$ in the plane at that time. Let the meeting point of the two robots be different from $S$, say at distance $x$ from $S$.
    \begin{figure}[!htb]
            \centering
            \includegraphics[width=4cm]{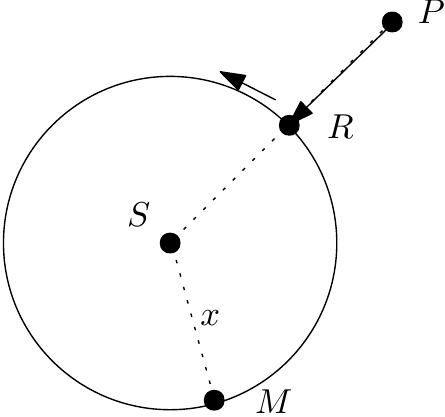}
            \caption{Meeting at a point other than $S$ in the {\tt NoAxis} model. Robot $r_1$ has speed $v_1$ and is located at $S$, while robot $r_2$ has speed $v_2$ and is located at $P$.}
            \label{fig:noaxis-meet}
        \end{figure}
    However, in the {\tt NoAxis} model the robots have no sense of direction and the robot $r_1$ has no way of knowing in what direction the robot $r_2$ is coming from. Therefore since the robot system is rotation invariant both robots must maintain their distance $x$ from the source $S$ while at the same time circulating the perimeter of the resulting curve so as to meet each other; clearly, in our case, this curve must be a circle of radius $x$.  
    As depicted in Figure~\ref{fig:noaxis-meet}, the robot $r_2$ reaches the curve at some point $R$ on the perimeter of the circle by heading towards $S$. 

    Assume the two robots meet at a point $M$ on the perimeter of the circle at distance $x$ from $S$. The adversary can arrange the directions of movement of $r_1, r_2$ so that the robots spend time at least $\frac{2\pi x-\epsilon}{v_2-v_1}$, where $\epsilon >0$ is arbitrarily small, in order to meet at the point $M$.  It follows that the two robots will meet at $M$ in total time at least $\frac{|PR|}{v_2} + \frac{2\pi x -\epsilon}{v_2-v_1}$ and it will take that plus at least $(D-x)/v_2$ time to deliver the bomb.
    The optimal time, though, is at most
    $$\frac{|PR| + x}{v_1+v_2} + \frac{D-x}{v_2}.$$
    It follows, then, that the competitive ratio is unbounded (as $v_2 - v_1 \rightarrow 0$)
    ~\qed
\end{proof}

\subsection{{\tt VisibleBoundary} Model}\label{sec:online_noaxis_visibleboundary}

First, we study the model where the critical distance $D$ is known.
Clearly, the optimally competitive algorithm for the {\tt OneAxis} is not feasible in this model, since robots cannot decide on a common axis or direction to move along and potentially meet for a handover.
By Lemma~\ref{lm:meet_at_source}, the robots only hope for cooperation is by meeting at $S$.
So if robots are going to collaborate, at least one of the robots will need to wait at $S$ for the other robot to arrive.
Clearly, it cannot do this forever though --- it may be the case that the other robot is so far away or slow that the first robot may as well have delivered the bomb by itself!
On the other hand, if the first robot simply commits to delivering the bomb by itself without waiting for the other robot to arrive, it may miss an opportunity to deliver the bomb very quickly if the other robot arrives at $S$ shortly after and is very fast.
It would seem, then, that an optimal algorithm must balance the cost of waiting for the other robot to arrive and completing the delivery without collaboration.

The main result of this section is the following theorem:
\begin{theoremrep}
    There exists an algorithm in the
     {\tt NoAxis}/{\tt VisibleBoundary} model with optimal competitive ratio $1 + \sqrt{2}$.
\end{theoremrep}
\begin{proof}
    The proof is a consequence of the following two Lemmas~\ref{lm:vb_lower},~\ref{lm:vb_upper}.
    ~\qed
\end{proof}

We'll start by proving that there exists no algorithm with a better competitive ratio:
\begin{lemmarep}\label{lm:vb_lower}
    Any online algorithm for the {\tt NoAxis}/{\tt VisibleBoundary} model has competitive ratio at least $1 + \sqrt{2}$.
\end{lemmarep}
\begin{proof}
    Consider the case where one robot starts with the bomb at the source ($d_1=0$) and has a speed of $v_1=1$.
    Suppose also there is another robot, but its distance $d_2$ and speed $v_2$ are unknown.
    First, observe that in this case there must exist an optimal algorithm of the following form: The first robot waits a time $t$ and, if the other robot has not yet arrived, takes the bomb to the boundary for delivery. Otherwise, the faster robot takes the bomb.
    This follows from the fact that the robots cannot meet at any point other than the source (Lemma~\ref{lm:meet_at_source}) and immediately leaving with the bomb for some finite time $t>0$ has arbitrarily large competitive ratio (by placing a fast robot with speed $v^\prime$ at a distance $d^\prime$ the competitive ratio is at least $t \cdot v^\prime/(d^\prime+D)$ which is arbitrarily large as $v^\prime \rightarrow \infty$ and $d^\prime \rightarrow 0$).
    
    We will show that for any $t$, we can choose a $v_2$ and $d_2$ which forces the competitive ratio to be at least $1 + \sqrt{2}$.
    The case where $t \leq \frac{1}{\sqrt{2}}$ is fairly simple.
    We can simply set $v_2=\frac{2}{\epsilon}$ and $d_2=1+\frac{2t}{\epsilon}$.
    Observe $d_2/v_2 = t + \epsilon/2$ (so robot $2$ arrives after the waiting period) and $(d_2 + 1)/v_2 = t + \epsilon$, so for any arbitrarily small $\epsilon > 0$, the second robot \textit{could have} delivered the bomb in time $t+\epsilon$.
    Then the competitive ratio is at least
    \begin{align*}
        \frac{t+1}{t+\epsilon} \geq \frac{1+\sqrt{2}}{1+\sqrt{2}\epsilon}
    \end{align*}
    which follows since the expression is decreasing with $t$ and $t > 1/\sqrt{2}$.
    Observe then, that the competitive ratio can be made arbitrarily close to $1+\sqrt{2}$ with some $\epsilon > 0$.

    The more interesting case is when $t > 1 / \sqrt{2}$.
    In this case, we set 
    \begin{align*}
        v_2 = \frac{\left(2+\sqrt{2}\right) t-\sqrt{2}}{2 \sqrt{2} t-2}
    \end{align*}
    and then 
    \begin{align*}
        d_2 = \frac{-t v^2-t v-v^2+\sqrt{2} v+\sqrt{2}+1}{-\sqrt{2} v-v+\sqrt{2}+1}.
    \end{align*}

    Observe $d_2$ and $v_2$ are defined and $d_2/v_2 > t$ for all $t > 1 / \sqrt{2}$, so the second robot always arrives after the waiting period (and thus does not participate).
    Finally, a little algebra and simplification shows the competitive ratio in this case is \textit{exactly} $1 + \sqrt{2}$:
    \begin{align*}
        \frac{t+1}{\frac{d_2}{v_2} + \left(1-\frac{d_2}{v_2}\right)/v_2} = 1+\sqrt{2}.
    \end{align*}
    ~\qed
\end{proof}

Now, we propose Algorithm~\ref{alg:online_noaxis_visibleboundary} and then prove its competitive ratio to be at most $1 + \sqrt{2}$.
\begin{algorithm}[!h]
    \caption{Online Algorithm for {\tt NoAxis} model for robot with speed $v$}\label{alg:online_noaxis_visibleboundary}
    \begin{algorithmic}[1]
        \State move to $S$ and wait for $D/v$ time
        \If {other robot arrives within $D/v$ time}
            \State faster of two robots picks up bomb and moves toward boundary for delivery
        \Else
            \State pick up bomb and move toward boundary for delivery
        \EndIf 
    \end{algorithmic}
\end{algorithm}

\begin{lemmarep}\label{lm:vb_upper}
    Algorithm~\ref{alg:online_noaxis_visibleboundary} has a competitive ratio of at most $1+\sqrt{2}$.
\end{lemmarep}
\begin{proof}
    Consider two robots with initial distances from the bomb $d_1$, $d_2$ and speeds $v_1$, $v_2$, respectively.
    Suppose, without loss of generality that $v_1 \leq v_2$ and that $D=1$.
    Throughout the proof, we'll refer to the robot with speed $v_1$ as ``the slow robot'' and the robot with speed $v_2$ as ``the fast robot''.
    For Algorithm~\ref{alg:online_noaxis_visibleboundary}, there are four interesting cases.

    \noindent\textbf{Case 1}: the slow robot arrives first and the fast robot does not arrive within the waiting period, or $\frac{d_1+1}{v_1}<\frac{d_2}{v_2}$.
    In this case, the delivery time of Algorithm~\ref{alg:online_noaxis_visibleboundary} is $(d_1+1)/v_2$. 
    The competitive ratio depends on the optimal delivery time, for which there are two possibilities.
    Either the slow robot and the fast robot cooperate to deliver the bomb or the fast robot is so far away that it does not participate in the optimal algorithm.
    In the latter situation, observe the competitive ratio is:
    \begin{align*}
        \frac{(d_1+2) / v_1}{(d_1+1) / v_1} = 1 + \frac{1}{d_1+1} \leq 2.
    \end{align*}
    In the former situation, however, the optimal delivery time is
    \begin{align}\label{eq:opt_del_time}
        \frac{d_1+d_2}{v_1+v_2} + \frac{1-\left(\frac{d_1+d_2}{v_1+v_2} v_1 - d_1\right)}{v_2} = \frac{1-d_2}{v_2} + \frac{2 (d_1+d_2)}{v_1+v_2}
    \end{align}
    and so the competitive ratio is    
    \begin{align}
        \frac{
            \frac{d_1+2}{v_1}
        }{
            \frac{1-d_2}{v_2} + \frac{2 (d_1+d_2)}{v_1+v_2}
        }
        &\leq \frac{2 d_2 (1 + d_2)}{1+d_2^2} \label{eq:vb_upper_1} \\ 
        &\leq 1 + \sqrt{2}. \label{eq:vb_upper_2}
    \end{align}

    Inequality~\eqref{eq:vb_upper_1} follows by Lemma~\ref{lm:start} (a maximum competitive ratio results in setting $d_1 = 0$) and by substituting $\frac{d_2v_1}{d_1+1}$ for $v_2$ (since $\frac{d_1+1}{v_1}<\frac{d_2}{v_2})$.
    Then~\eqref{eq:vb_upper_2} follows directly since the expression is maximized when $d_2 = 1+\sqrt{2}$.
  
    \noindent\textbf{Case 2}: the fast robot arrives first and the slow robot does not arrive within the waiting period, or $\frac{d_2+1}{v_2}<\frac{d_1}{v_1}$.
    In this case, the delivery time of Algorithm~\ref{alg:online_noaxis_visibleboundary} is $(d_2 + 2) / v_2$ while the optimal delivery time is at least $(d_2 + 1) / v_2$ since the fast robot arrives at the bomb first and would deliver it without collaboration. Thus, the competitive ratio for this case is at most $2$:
    \begin{align*}
        \frac{(d_2+2) / v_2}{(d_2+1) / v_2} = 1 + \frac{1}{d_2+1} \leq 2.
    \end{align*}

    \noindent\textbf{Case 3}: the fast robot arrives first and the slow robot arrives during the waiting period --- $\frac{d_2}{v_2}<\frac{d_1}{v_1}\leq\frac{d_2+1}{v_2}$). 
    Similarly, the delivery time of Algorithm~\ref{alg:online_noaxis_visibleboundary} in this case is $(d_1 + 2) / v_1$ while the optimal delivery time is at least $(d_1 + 1) / v_1$ since the fast robot is so far away that it cannot arrive within a distance of $1$ from the bomb in that time. Thus the competitive ratio for this case is also at most $2$:
    \begin{align*}
        \frac{(d_1+2) / v_1}{(d_1+1) / v_1} = 1 + \frac{1}{d_1+1} \leq 2.
    \end{align*}

    \noindent\textbf{Case 4}: the slow robot arrives first and the fast robot arrives during the waiting period: $\frac{d_1}{v_1}<\frac{d_2}{v_2}\leq\frac{d_1+1}{v_1}$. In this case, an optimal-time delivery must involve cooperation between the two robots, and thus is given by~\eqref{eq:opt_del_time}.
    The competitive ratio can then be written
    \begin{align*}
        \frac{
            \frac{d_2+1}{v_2}
        }{
            \frac{1-d_2}{v_2} + \frac{2 (d_1+d_2)}{v_1 + v_2}
        } \leq \frac{
            \frac{d_2+1}{v_2}
        }{
            \frac{1-d_2}{v_2} + \frac{2 d_2}{v_1 + v_2}
        }
        \leq 1 + \frac{2d_2}{1 + d_2^2} \leq 2
    \end{align*}
    which follows since, by Lemma~\ref{lm:start}, setting $d_1=0$ cannot decrease the competitive ratio (first inequality) and then by substituting $\frac{v_2}{d_2}$ for $v_1$ (since $\frac{d_2}{v_2}\leq\frac{d_1+1}{v_1})$ and simplifying (second inequality).
    Finally, the last inequality follows since $1+\frac{2 d_2}{1 + d_2^2}$ has a maximum value of $2$ (at $d_2 = 1$).
    ~\qed    
\end{proof}

\subsection{{\tt DiscoverableBoundary} Model}
\label{sec:Online Algorithm for the DiscoverableBoundary Model}

In this section, we consider the scenario where the robots are disoriented ({\tt NoAxis} model) and do not know the distance $D$ of the boundary from the source, but {\em can} discover it by passing through a point on the boundary or by encountering another robot which has previously discovered it.

\begin{theoremrep}
    Any algorithm for the {\tt NoAxis / DiscoverableBoundary} model has competitive ratio at least $3$. 
\end{theoremrep}

\begin{proof}
    Suppose a robot $r_1$ starts at the source and the other robot at some initial position not at the source.
    Then the robot, as its first step in an algorithm, can do one of three things:
    \begin{enumerate}
        \item Wait for some time $t > 0$
        \item Leave the source with the bomb
        \item Leave the source without the bomb
    \end{enumerate}
    We will show for each of these cases that there is a configuration that results in a competitive ratio of at least $3$.

    First, let's consider Case 1 (the robot waits for some time $t > 0$). In this case, we let $r_1$ be the fast robot with speed $v_1 = 1$, set the distance of the boundary $D = t/3$, and position the other robot so that it does not arrive in time $t$ to the source. Observe in this case the fast robot could have delivered the bomb by itself in time $t/3$ but instead takes time at least $t$.
    
    For Case 2 (the robot leaves the source with the bomb), we simply set $r_1$ to be the \textit{slow} robot and set the fast robot so close that it arrives just after the $r_1$ leaves with the bomb.
    For example, suppose $r_1$ leaves the source and travels a maximum distance $x$ from the source with the bomb.
    Then, we let $D = x$ and $v_1 = 1$ (so that the slow robot delivers the bomb at time $1$) , $d_2=1/2$, $v_2 = \frac{9}{2D}$ then observe the fast robot could have delivered the bomb in time $D/3$.

    For Case 3 (the robot leaves the source without the bomb), we consider two subcases.
    First, if the robot never returns to the source, then clearly the algorithm is unbounded because by setting $v_2 = \epsilon$ for some arbitrarily small $\epsilon > 0$, the bomb will take an arbitrarily long amount of time to be delivered to the destination.
    That leaves the case where the robot leaves the source and returns after traveling a maximum distance $x$ from the source.
    In this case, we can simply set $r_1$ to be the fast robot with speed $v_1=1$ and $r_2$ to be a useless slow robot with speed $v_2=\epsilon$ for some arbitrarily small $\epsilon > 0$.
    Then by setting $D=x$, the robot takes at least time $3D$ to deliver the bomb ($2x = 2D$ time to travel $x$ and back to the source and another $D$ time to deliver the bomb) while the optimal offline algorithm clearly would take just $D$ time.
    ~\qed
\end{proof}

Now, we present Algorithm~\ref{alg:online_noaxis_discoverableboundary} with competitive ratio of $15/4$.
\vspace*{-0.5cm}
\begin{algorithm}[!h]
    \caption{Online Algorithm for {\tt DiscoverableBoundary} Model for robot with speed $v$}\label{alg:online_noaxis_discoverableboundary}
    \begin{algorithmic}[1]
    \State move to $S$
    \If {discovered boundary on the way to $S$}
        \State wait time $D/v$;
    \Else 
        \State move away from $S$ (without the bomb) until arriving at the boundary
        \State return to $S$
    \EndIf
    \If {bomb is still at $S$}
        \State take bomb to the boundary
    \EndIf
    \end{algorithmic}
\end{algorithm}
\vspace*{-0.5cm}

\begin{theoremrep}
    Algorithm~\ref{alg:online_noaxis_discoverableboundary} has a competitive ratio of exactly $15/4$.
\end{theoremrep}
\begin{proof}
    For the lower bound, let $D=1$, $d_1=0$, $v_1=2/(3-2\epsilon)$, $d_2=1-\epsilon$, and $v_2=1$ for some arbitrarily small $\epsilon > 0$.
    Observe $r_1$ delivers the bomb in this case since 
    $2/v_1 < (2+d_2)/v_2 \Leftrightarrow 3-2\epsilon < 3-\epsilon$.
    Thus, the competitive ratio can be written:
    \begin{align*}
        \frac{
            \frac{3}{v_1}
        }{
            \frac{1}{1 + v_1} + 1 - \frac{1}{1 + v_1} \cdot v_1
        } &= \frac{3(1+v_1)}{2v_1} 
        = 3 + \frac{9 - 18 \epsilon}{12 - 10 \epsilon + 4 \epsilon^2}
    \end{align*}
    whose limit is $15/4$ as $\epsilon \rightarrow 0$.

    For the upper bound, observe that if the faster robot arrives at $S$ first, then the optimal offline delivery time is $(1+d_1)/v_1$ and there are three possible outcomes for the online algorithm.
    First, if the fast robot started outside the critical zone ($d_2 > 1$), then the delivery time is $(2+d_2)/v_2$ and the competitive ratio is
    $$\frac{(2+d_2)/v_2}{(1+d_2)/v_2} \leq 2.$$
    If the fast robot started {\em inside} the critical zone, however, then it is possible the slow robot takes the bomb before the fast robot returns from searching for the boundary $\left( (d_1 + 1)/v_1 < (d_2 + 2)/v_2 \right)$ (Clearly this can only occur if the slow robot starts outside the critical zone, and thus discovers the boundary on its way to $S$).
    In this case, since $v_1 > v2 (d_1+1)/(d2+2)$, the competitive ratio is
    $$
    \frac{(d_1+2)/v_1}{(d_2+1)/v_2} 
    = \frac{v_2}{v_1} \frac{d_1+2}{d_1+1} 
    \leq \frac{(2+d_1)(2+d_2)}{(1+d_1)(1+d_2)}.
    $$
    Observe the expression clearly increases w.r.t. $d_0$ and decreases w.r.t. $d_1$ and thus is maximized when $d_1=1$ and $d_2=0$, yielding an upper bound of $3$ on the competitive ratio in this case.
    Finally, if the slow robot does not arrive in time to take the bomb, the competitive ratio is 
    $$\frac{(d_2+3)/v_2}{(d_2+1)/v_2} \leq 3.$$
    Thus, if the faster robot arrives at $S$ first, the competitive ratio of Algorithm~\ref{alg:online_noaxis_discoverableboundary} is at most $3$.

    Now we focus our attention on the slightly more involved scenario where the slow robot arrives at $S$ first.
    Clearly if in an optimal offline algorithm, the slow robot delivers the bomb by itself (when $(d_1+1)/v_1 \leq (d_2-1)/v_2$), then Algorithm~\ref{alg:online_noaxis_discoverableboundary} is also optimal.
    Thus, we may assume $(d_1+1)/v_1 > (d_2-1)/v_2$.
    Furthermore, since we assume the slow robot arrives at $S$ first, a maximum competitive ratio must exist when $d_1=0$ (by Lemma~\ref{lm:start}).
    Considering these assumptions, the optimal offline delivery time, then, is
    $$\frac{d_2}{v_1+v_2}+\left( 1 - \frac{d_2}{v_1+v_2} v_1 \right) / v_2 = \frac{1-d_2}{v_2} + \frac{2 d_2}{v_1+v_2}.$$
    There are three cases to consider.

    \noindent {\bf Case 1}: The fast robot starts outside the critical zone $d_2 \geq 1$ and the slow robot delivers the bomb $2/v_1 < (d_2+1)/v_2$.
    In this case, the competitive ratio is
    $$\frac{3/v_1}{\frac{1-d_2}{v_2} + \frac{2 d_2}{v_1+v_2}} < \frac{3}{2} + \frac{6 d_2}{3+d_2^2} < \frac{3+2\sqrt{3}}{2}$$
    since $v_2 < v_1 (d_2+1)/2$ and the left-hand side of the first inequality above is increasing w.r.t $v_2$.
    The second inequality follows since the middle expression is maximized at $d_2=\sqrt{3}$.

    \noindent {\bf Case 2}: The fast robot starts outside the critical zone $d_2 \geq 1$ and the fast robot delivers the bomb $2/v_1 \geq d_2/v_2$.
    In this case, the competitive ratio is
    \begin{align*}
        \frac{
            \min \left[ (d_2+1)/v_2, 1/v_1 \right] + 1/v_2
        }{
            \frac{1-d_2}{v_2} + \frac{2 d_2}{v_1+v_2}
        }.
    \end{align*}
    First, suppose $(d_2+1)/v_2 \leq 1/v_1$, then the competitive ratio is
    \begin{align*}
        \frac{
            (d_2+2)/v_2
        }{
            \frac{1-d_2}{v_2} + \frac{2 d_2}{v_1+v_2}
        } \leq \frac{(d_2 + 2)^2}{2+d_2+d_2^2} \leq \frac{16}{7}
    \end{align*}
    since the middle expression is maximized at $d_2 = 2/3$.
    On the other hand, if $(d_2+1)/v_2 > 1/v_1$, then the competitive ratio is
    \begin{align}
        \frac{
            1/v_1 + 1/v_2
        }{
            \frac{1-d_2}{v_2} + \frac{2 d_2}{v_1+v_2}
        } 
        &\leq \frac{(2+d_2)(v_1+v_2)}{v_1-d_2 v_1 + v_2 + d_2 v_2} 
        = \frac{v_1 (2+d_2) + v_2 (2+d_2)}{v_1(1-d_2) + v_2(1+d_2)} \label{eq:discover_upper_case2_1}\\
        &\leq \frac{(2+d_2)^2}{2+d_2+d_2^2} \label{eq:discover_upper_case2_2}\\
        &\leq \frac{9}{4}. \label{eq:discover_upper_case2_3}
    \end{align}
    Inequality~\eqref{eq:discover_upper_case2_1} follows by substituting $v_1$ in the numerator with $v_2 / (d_2 + 1)$ (since $v_1 > v_2 / (d_2 + 1)$).
    After simplifying, it's clear that Inequality~\eqref{eq:discover_upper_case2_2} follows for the same reason.
    Then Inequality~\eqref{eq:discover_upper_case2_3} follows since the right-hand side of~\eqref{eq:discover_upper_case2_2} is maximized (on $d_2 \geq 1$) at $d_2=1$.

    \noindent {\bf Case 3}: The fast robot starts inside the critical zone $d_2 < 1$ and the slow robot delivers the bomb $2/v_1 < (d_2+2)/v_2$.
    In this case, observe the competitive ratio is
    $$\frac{3/v_1}{\frac{1-d_2}{v_2} + \frac{2 d_2}{v_1+v_2}} < \frac{3 (2+d_2)(4+d_2)}{2(4+d_2+d_2^2)} < \frac{15}{4}$$
    since $v_2 < v_1 (d_2+2)/2$ and the left-hand side of the first inequality is increasing w.r.t. $v_2$.
    The second inequality follows since the middle expression is maximized at $d_2=1$.

    \noindent {\bf Case 4}: The fast robot starts inside the critical zone $d_2 < 1$ and the fast robot delivers the bomb $2/v_1 \geq (d_2+2)/v_2$.
    In this case, the competitive ratio is
    \begin{align*}
        \frac{(d_2+3)/v_2}{\frac{1-d_2}{v_2} + \frac{2 d_2}{v_1+v_2}} 
        \leq \frac{(3+d_2)(4+d_2)}{4+d_2+d_2^2}
        \leq \frac{5}{3} + \frac{8 \sqrt{2/5}}{3}.
    \end{align*}
    The first inequality above follows since $v_1 \leq 2 v_2 / (d_2+2)$ and the second inequality follows since the middle expression is maximized at $d_2=2/3 \cdot \left( \sqrt{10} - 2 \right)$.
    
    Case 3's competitive ratio of $15/4$ dominates all other cases and is exactly the situation used to show the lower bound for the algorithm above.
    ~\qed
\end{proof}

\subsection{{\tt InvisibleBoundary} Model}
\label{sec:online_noaxis_invisibleboundary}

Finally, we analyze an online algorithm under much stricter conditions.
Robots cannot perceive the boundary at any time and therefore can never know the critical distance.
It follows, then, that any valid algorithm must involve robots carrying the bomb away from the source without knowing how far they must take it in order to terminate.

Assume the slow robot starts at the source. By Lemma~\ref{lm:meet_at_source} there is no online algorithm with bounded competitive ratio unless the two robots have a meeting at the source. Further, it is easy to see that if the slow robot leaves the source without the bomb then unless it returns to the source there can be no online algorithm with bounded competitive ratio.

 \begin{lemmarep}\label{lm:need_lower_bound_D}
    There exists no algorithm with constant competitive ratio for any instance of the problem under the {\tt NoAxis}/{\tt InvisibleBoundary} model in which one robot starts at $S$ and no lower bound on $D$ is known to the robots.
 \end{lemmarep}
 \begin{proof}
    Suppose robot $r_1$ with speed $v_1$ is initially placed at the source. 
    Then it may do one of three things:
    \begin{enumerate}
        \item wait at $S$ for a time $t$
        \item leave $S$ for a time $t$ without the bomb
        \item leave $S$ for a time $t$ with the bomb
    \end{enumerate}

    In the first and second cases, we simply set $D$ to be so small that the robot could have delivered the bomb to the boundary in an arbitrarily small fraction of the time it waited or left $S$ for.
    The competitive ratio is at least $\frac{D/v_1 + t}{D/v_1} = 1 + \frac{v_1 t}{D}$ which approaches infinity as $D \rightarrow 0$.
    In the third case, we set $D = t$ and $v_1=1$ and configure a second, much faster robot to arrive at $S$ just after $r_1$ leaves with the bomb.
    By Lemma~\ref{lm:meet_at_source}, the robots must meet at the source, otherwise, the algorithm has unbounded competitive ratio.
    Thus, the robots cannot meet before time $t$ and the competitive ratio is at least $\frac{t}{(d_2 + D) / v_2} = \frac{v_2 t}{d_2 + D}$ since $t$ is the earliest time the two robots can meet.
    Again, this approaches infinity as $v_2 \rightarrow \infty$ for any constant $d_2$.
    ~\qed
 \end{proof}

In order to provide an online algorithm with constant competitive ratio, we make the necessary assumption (by Lemma~\ref{lm:need_lower_bound_D}) that the critical distance $D \geq 1$.
In the sequel, we provide an algorithm that involves the first robot arriving at $S$ taking the bomb a certain distance away from $S$ and then returning (without the bomb) to see if a faster robot has arrived. If a faster robot \textit{has} arrived, it shares information about the distance and direction of the bomb and allows the faster robot to complete the delivery. Otherwise, it travels back to where it left the bomb and carries it a bit further, expanding the distance each time. Formally, we present Algorithm~\ref{alg:online_noaxis_invisibleboundary} below.

We now prove a theorem that gives an upper bound on the competitive ratio of Algorithm~\ref{alg:online_noaxis_invisibleboundary}. 
Note that Algorithm~\ref{alg:online_noaxis_invisibleboundary} uses the as yet unspecified expansion factor $a>1$. 
The optimal selection of $a$ will turn out to be $a=\frac{3+\sqrt{17}}4$ and this will be determined in the course of the proof of the following theorem. 

\begin{theoremrep}\label{hm:zigzag}
    For two robots, Algorithm~\ref{alg:online_noaxis_invisibleboundary} with $a=\frac{3+\sqrt{17}}4$ delivers the bomb to the boundary in at most $\frac{7+\sqrt{17}}2$ times the optimal offline time.
\end{theoremrep}
\begin{proof}
    Let $D\geq 1$ be the critical distance (distance from $S$ to the boundary) which is at least the common unit distance. 
    Consider two robots $r_1, r_2$ with respective speeds $v_1 < v_2$ and  let $d_i$ be the distance of the initial position of robot $r_i$ from the source $S$, for $i=1,2$. 

    First, we consider the case where the faster robot when moving at full speed cannot arrive at $S$ before the slow robot does. In view of Lemma~\ref{lm:start}, without loss of generality we may assume the slow robot $r_1$ starts at the source $S$, i.e., $d_1=0$, and $d_2 $ is the distance of robot $r_2$ from the source at that time. 
    Consider Algorithm~\ref{alg:online_noaxis_invisibleboundary}. We consider two main cases depending on whether or not the faster robot $r_2$ arrives at $S$ during robot $r_1$'s waiting period.

    \noindent {\bf Case 1.} Robot $r_2$ arrives during robot $r_1$'s initial waiting period.\\

    In this case $r_2$, after comparing its speed with $r_1$, will deliver the bomb. 
    Observe that in this case $\frac{d_2}{v_2} \leq \frac 2{v_1}$, i.e., $d_2 \leq 2 \frac{v_2}{v_1}$. 
    To compute the resulting competitive ratio we must take into account whether or not the two robots meet inside the circle centered at $S$ and radius $D$. 
    In the optimal algorithm if we assume the robots are moving towards each other at full speed they will meet at time $\frac{d_2}{v_1+v_2}$. 

    If $\frac{d_2 v_1}{v_1+v_2} \geq D$ then in the optimal algorithm robot $r_1$ will deliver the bomb to the perimeter. Since $d_2 \leq 2 \frac{v_2}{v_1}$ we conclude that
    $$
    CR \leq \frac{\frac{d_2 +D}{v_2}}{\frac{D}{v_1}} \leq \frac{v_1}{v_2} + \frac{2}{D} \leq 3 .
    $$
    If $\frac{d_2 v_1}{v_1+v_2} < D$ then in the optimal algorithm the two robots will meet, robot $r_1$ will hand over the bomb to $r_2$ which in turn will deliver it to the perimeter. Therefore we have that
    \begin{align*}
    CR &\leq 
    \frac{\frac{d_2 +D}{v_2}}{\frac{d_2}{v_1+v_2} + \frac{D- \frac{d_2v_1}{v_1+v_2}}{v_2}} = \frac{\frac{d_2 +D}{v_2}}{\frac{d_2}{v_1+v_2} \left( 1-\frac{v_1}{v_2}\right) + \frac{D}{v_2}} \\
    & \leq \frac{\frac{d_2 +D}{v_2}}{\frac{d_2}{2v_2} \left( 1-\frac{v_1}{v_2}\right) + \frac{D}{v_2}} = \frac{d_2+D}{\frac{d_2}{2} \left( 1-\frac{v_1}{v_2}\right) + D}
    \end{align*}
    Finally, we show that the right-hand side above is at most $4$. Indeed, this is equivalent to showing that $d_2+D \leq 4 \frac{d_2}{2} \left( 1-\frac{v_1}{v_2}\right) + 4 D$. This simplifies to the equivalent inequality 
    \begin{equation}
    \label{eq:maincase1}
    1 \leq 2 \left( 1-\frac{v_1}{v_2}\right) + 3 \frac D{d_2}. 
    \end{equation}
    Using the fact that $d_2 \leq 2 \frac{v_2}{v_1}$ we observe that 
    \begin{equation}
    \label{eq:itisok}
    2 \left( 1-\frac{v_1}{v_2}\right) + 3 \frac D{d_2}
    \geq 
    2 - \frac{4}{d_2} + 3 \frac D{d_2} 
    \geq 
    1 + \frac{6D -4 }{ d_2} 
    \geq 1,
    \end{equation}
    where Inequality~\eqref{eq:itisok} is valid since by assumption $D\geq 1$. This confirms Inequality~\eqref{eq:maincase1} and completes the proof for the case when robot $r_2$ arrives during robot $r_1$'s initial waiting period.

    \noindent {\bf Case 2.} Robot $r_2$ arrives after robot $r_1$ has completed its initial waiting period, i.e. $\frac{d_2}{v_2} > \frac 2{v_1}$.\\
    In the case where robot $r_2$ does not arrive at $S$ before robot $r_1$ has completed its last visitation of the source $S$ during its zigzag trajectory then $r_1$ will deliver the bomb. Consider the last round $k$ such that $a^k < D$. It follows that $a^k < D \leq a^{k+1}$. Clearly, the total distance covered by the robot $r_1$ until the boundary was reached is equal to
    $$
    2 \left( a^0 + a^1 + \cdots + a^k \right) + D = 2 \frac{a^{k+1} -1}{a-1} + D  .
    $$
    It follows that this is traversed in total time 
    \begin{equation}
    \label{eq:zigzagbef}
    \frac{2 \frac{a^{k+1} -1}{a-1} + D}{v_1} < \frac{2 \frac{a^{k+1}}{a-1} + D}{v_1} < \frac{2 \frac{a D}{a-1} + D}{v_1}
    \end{equation}
    The resulting competitive ratio satisfies
    \begin{equation}
    \label{eq:zigzag}
    CR \leq \frac{ \frac{2 \frac{a D}{a-1} + D}{v_1} }{\frac D{v_1}} = \frac{2a}{a-1} + 1
    \end{equation}

    Now consider the case where robot $r_2$ does arrive at $S$ before robot $r_1$ has reached the boundary. This also means $r_2$ arrives at $S$ before robot $r_1$ has completed its last visitation of the source $S$ during its zigzag trajectory. Therefore since the robot $r_2$ is at distance $d_2$ from $S$, for some round $i$ we must have that
    \begin{equation}
    \label{eq:zigzagbef1}
    \frac 2{v_1} \left( a^0 + a^1 + \cdots + a^{i-1} \right)
    < \frac{d_2}{v_2} \leq 
    \frac 2{v_1} \left( a^0 + a^1 + \cdots + a^{i} \right),
    \end{equation}
    which is readily simplified to
    \begin{equation}
    \label{eq:1zigzag4}
    \frac 2{v_1} \cdot \frac{ a^i-1}{a-1}
    < \frac{d_2}{v_2} \leq 
    \frac 2{v_1} \cdot \frac{ a^{i+1}-1}{a-1} .
    \end{equation}
    Since the bomb is not at $S$, the robot $r_2$ must wait until robot $r_1$ arrives at time $\frac 2{v_1} \frac{ a^{i+1}-1}{a-1}$ to learn its location. 
    By exchanging speeds, robot $r_1$ now knows that $r_2$ is the faster robot and shares the direction of the bomb with it which is then delivered to the boundary in additional time $\frac D{v_2}$ (the fast robot moves to the bomb, picks it up, and continues moving away from $S$). Also, observe that 
    \begin{equation}
    \label{eq:zigzag1}
    \frac{a^i}{v_1} < D
    \end{equation}
    since $r_1$ could not reach the boundary during the $i$th round. It follows that the total time required for the bomb to be delivered to the boundary using Algorithm~\ref{alg:online_noaxis_invisibleboundary} is at most 
    \begin{equation}
    \label{eq:zigzag2}
    \frac 2{v_1} \cdot \frac{ a^{i+1}-1}{a-1} + \frac D{v_2} .
    \end{equation}

    Next, we consider the running time $T_{opt}$ of the optimal algorithm. Observe that the robots could meet in time $\frac{d_2}{v_1+v_2}$ because at the start robot $r_1$ is at the source $S$ and hence they are at distance $d_2$.  Therefore
    \begin{enumerate}
    \item \label{item1}
    if $\frac{d_2}{v_1+v_2} > \frac D{v_1}$ then $T_{opt} = \frac D{v_1}$, and
    \item \label{item2}
    if $\frac{d_2}{v_1+v_2} \leq \frac D{v_1}$ then $T_{opt} = \frac{d_2}{v_1+v_2} + \frac{D - \frac{d_2 v_1 }{v_1+v_2}}{v_2}$.
    \end{enumerate}

    If $\frac{d_2}{v_1+v_2} > \frac D{v_1}$ then 
    by assertion~\ref{item1} above,  $T_{opt} = \frac D{v_1}$ and hence the competitive ratio of Algorithm~\ref{alg:online_noaxis_invisibleboundary} is easily seen to satisfy
    \begin{equation}
    \label{eq:final1}
    CR \leq \frac{\frac 2{v_1} \cdot \frac{ a^{i+1}-1}{a-1} + 
    \frac D{v_2}}{\frac D{v_1}} \leq \frac{2	a}{a-1}+ \frac{v_1}{v_2} \leq \frac{2	a}{a-1} + 1,
    \end{equation}
    since $v_1 \leq v_2$.

    So assume that $\frac{d_2}{v_1+v_2} \leq \frac D{v_1}$ in which case assertion~\ref{item2} above is valid and the optimal time is given by the formula
    $$
    T_{opt} = \frac{d_2}{v_1+v_2} + \frac{D - \frac{d_2 v_1 }{v_1+v_2}}{v_2}
    $$
    and the competitive ratio of Algorithm~\ref{alg:online_noaxis_invisibleboundary} satisfies
    \begin{equation}
    \label{eq:zigzag3}
    CR \leq \frac{\frac 2{v_1} \cdot \frac{ a^{i+1}-1}{a-1} + 
    \frac D{v_2}}{ \frac{d_2}{v_1+v_2} + \frac{D - \frac{d_2 v_1 }{v_1+v_2}}{v_2} } .
    \end{equation}

    Next, we consider two cases depending on whether or not robot $r_2$ starts inside or outside the boundary.

    {\em Case 1.} Assume $d_2 \leq D$, namely robot $r_2$ starts inside the boundary.\\ 
    By Inequality~\eqref{eq:zigzag3} we have
    \begin{align}
    CR & \notag \leq 
    \frac{\frac 2{v_1}  \cdot \left( \frac{ a^{i+1}-a}{a-1} + 1 \right) + 
    \frac D{v_2}}{ \frac{d_2}{v_1+v_2} + \frac{D - \frac{d_2 v_1 }{v_1+v_2}}{v_2} } \\
    &< \notag \frac{ \frac{d_2}{v_2} a  + \frac 2{v_1}  + \frac D {v_2} }{ \frac{d_2}{v_1+v_2} \left( 1 - \frac 1{v_2} \right)+ \frac{D}{v_2} } \mbox{~~~(by Inequality~\eqref{eq:1zigzag4}, $\frac 2{v_1} \frac{a^i-1}{a-1} < \frac{d_2}{v_2}$)}\\
    &\leq \notag \frac{ \frac{d_2}{v_2} a  + \frac 2{v_1}  + \frac D {v_2} }{ \frac{d_2}{2v_2} \left( 1 - \frac 1{v_2} \right)+ \frac{D}{v_2} } \mbox{~~~(since $v_1 \leq v_2$)} \\
    &\label{eq:final2} \leq a+3 .
    \end{align}
    To prove the last Inequality~\eqref{eq:final2}, multiply out and observe
    $$
    \frac{d_2}{v_2} a  + \frac 2{v_1}  + \frac D {v_2} \leq
    (a+3) \frac{d_2}{2v_2} \left( 1 - \frac 1{v_2} \right) + (a+3) \frac{D}{v_2} ,
    $$ 
    which is easy to prove since $\frac 2{v_1} < \frac{2D}{v_2}$, by Inequality~\eqref{eq:1zigzag4}, and by assumption $d_2 \leq D$. 

    {\em Case 2.} Assume $d_2 > D$, namely robot $r_2$ starts outside the boundary.\\
    Recall that we have that $\frac{d_2}{v_1+v_2} \leq \frac D{v_1}$. By Inequality~\eqref{eq:zigzag3} we have
    \begin{align}
    CR &\leq \notag
    \frac{\frac 2{v_1}  \cdot \frac{ a^{i+1}-1}{a-1}  + 
    \frac D{v_2}}{ \frac{d_2}{v_1+v_2} \left( 1 - \frac{v_1}{v_2} \right) + \frac{D}{v_2} }\\
    &< \notag
    \frac{ \frac{d_2a}{v_2}  + 
    \frac {d_2}{v_2}}{ \frac{d_2}{v_1+v_2} \left( 1 - \frac{v_1}{v_2} \right) + \frac{d_2}{v_1+v_2} \cdot \frac{v_1}{v_2} } 
    \mbox { ~~~(using Inequality~\eqref{eq:1zigzag4} and $D < d_2$)}\\
    &= \notag \frac{ \frac{a}{v_2}  + 
    \frac {1}{v_2}}{ \frac{1}{v_1+v_2} } \mbox{~~~(after dividing by $d_2$ and simplifying)}\\
    &\leq \label{eq:final3} 2a+2 \mbox{~~~(since $v_1 \leq v_2$)}
    \end{align}

    If we consider Inequalities~\eqref{eq:final2},~and~\eqref{eq:final3} we see that $a+3 \leq 2a + 2$ and therefore using Inequality~\eqref{eq:final1} we conclude that 
    $$
    CR \leq \max\left\{ \frac{2a}{a-1} + 1 , 2a+2 \right\} .
    $$
    Clearly, the upper bound above is minimized when $ \frac{2a}{a-1} + 1 = 2a+2$ since $\frac{2a}{a-1}+1$ is decreasing on $a \geq 1$ and $2a+2$ is increasing on $a \geq 1$.
    If we solve the resulting quadratic for $a$ we derive that $a = \frac{3+\sqrt{17}}4$ and therefore $CR \leq \frac{7+\sqrt{17}}2$.

    We now consider the case where the faster robot when moving at full speed arrives at $S$ before the slow robot does. We will prove that the bound $CR \leq \frac{7+\sqrt{17}}2$ on the competitive ratio is valid in this case as well. 

    According to Algorithm~\ref{alg:online_noaxis_invisibleboundary}, there are two cases to consider: 1) The slow robot arrives at $S$ during the faster robot's waiting period, and 2)  the slow robot arrives at $S$ after the faster has completed its initial waiting period.

    In the first case, we have that $\frac {d_1}{v_1} \leq \frac {2}{v_2}$ and the faster robot can take the bomb and move away from $S$ (toward the boundary). 
    Therefore the competitive ratio satisfies
    $$
    \frac{\frac{D+d_2+2}{v_2}}{\frac{D+d_2}{v_2}} \leq 1 + \frac{2}{D+d_2} \leq 3,
    $$
    where the last inequality is valid because $D\geq 1$.

    In the second case, we have that $\frac {d_1}{v_1} > \frac {2}{v_2}$ in which case the faster robot $r_2$ will pick up the bomb and execute a zigzag algorithm with expansion factor $a > 1$. Now the proof is similar to the proof for the slow robot in Inequality~\eqref{eq:1zigzag4}, namely, we obtain the following inequality
    \begin{equation}
    \label{eq:1zigzag4fast}
    \frac 2{v_2} \cdot \frac{ a^i-1}{a-1}
    < \frac{d_1}{v_1} \leq 
    \frac 2{v_2} \cdot \frac{ a^{i+1}-1}{a-1} .
    \end{equation}
    This also means that the two robots will meet and robot $r_2$, upon learning it is the faster robot, will move away from $S$ forevermore without turning back.
    The resulting competitive ratio satisfies
    \begin{equation}
    \label{eq:1zigzag4fast1}
    CR \leq \frac{\frac{\frac{2 a D}{a-1}+D}{v_2}}{\frac{d_2+D}{v_2}} \leq 
    \frac{\frac{2 a D}{a-1}+D}{d_2+D} \leq \frac{2a}{a-1}+1 .
    \end{equation}
    It is now easily seen that for $a = \frac{3+\sqrt{17}}4$ the right-hand side of the previous Inequality~\eqref{eq:1zigzag4fast1} is at most $\frac{7+\sqrt{17}}2$.
    The proof of Theorem~\ref{hm:zigzag} is now complete. 
    ~\qed
\end{proof}

Before proceeding to show a lower bound of $2+\sqrt{5}$ for any online algorithm for the {\tt NoAxis/InvisibleBoundary} model, we introduce a few basic concepts and ideas. 
Assume two robots $r_1, r_2$ with $v_1 = 1 < v_2$ and a source $S$. 
Recall that each robot knows the location of the source and its own location and speed but not the speed and location of the other robot. 
If the two robots meet, they can exchange information and determine which of the two is faster. 
If a robot knows it is faster than the other robot then if/when it acquires the bomb, it should simply move away from $S$ forever to guarantee eventual delivery. 
If a robot holds the bomb and is ``searching'' for the perimeter but does not know whether it is the faster robot then it must return to the source (without the bomb) to check whether or not the other robot is waiting there.
If it is, then it will share the direction of the bomb so that the fast robot can complete the delivery. 
Therefore the slow robot is forced to execute a zigzag strategy as defined below, otherwise, the adversary will make the competitive ratio arbitrarily large. 

\begin{algorithm}[!h]
    \caption{Online Algorithm for the {\tt InvisibleBoundary} Model for a robot speed $v$ and expansion factor $a$}\label{alg:online_noaxis_invisibleboundary}
    \begin{algorithmic}[1]
        \State $i \gets 0$
        \While {never encountered another robot}
            \State {move to $S$}
            \If {faster robot is at $S$}
                \State {share direction of bomb with faster robot and stay at $S$ forever}
            \ElsIf {slower robot is at $S$}
                \State {get direction of bomb from slow robot (if not already known)}
                \State {move to bomb, pick it up, and continue moving away from $S$ forever}
            \ElsIf{bomb is at $S$}
                \State {Wait for another robot for at most time $2/v$}
                \If {faster robot has not arrived}
                    \State {pick up bomb}
                    \State {move away from $S$ for a distance $a$}
                    \State {set the bomb down}
                \Else
                    \State stay at $S$ forever
                \EndIf
            \ElsIf{bomb is not at $S$ but its location is known}
                \State {move toward the bomb a distance of $a^i$ distance away from $S$}
                \State {pick up bomb}
                \State {move another $a^i$ distance away from $S$}
                \State {set the bomb down, marking its location}
            \Else
                \State {wait for other robot to return}
                \If {other robot is slower}
                    \State get direction of the bomb from other robot
                \EndIf
            \EndIf
            \State $i \gets i + 1$
        \EndWhile
    \end{algorithmic}
\end{algorithm}

A general algorithm is encapsulated by a strategy in which the robot starts at the source and executes Algorithm~\ref{alg:line2} implying a search at a distance $x_k$ in the $k$-th round of the algorithm, for each $k \geq 1$. 
The algorithm is parameterized by an infinite ordered sequence of positive distances $X=\{ x_1, x_2, \ldots, x_k, \ldots \}$ measured from the source that specifies the turning points that a moving robot will make. 
\begin{algorithm}[!h]
\caption{Zig-Zag Delivery Algorithm $(X)$}\label{alg:line2}
    \begin{algorithmic}[1]
        \State {{\bf Input:} Infinite sequence $X=\{x_1, x_2, \ldots,x_k,\ldots\}$ with $0 < x_k < x_{k+1}$  for all $k \geq 1$;}
        \For{$k \leftarrow 1,2,3,\ldots$}
        \If {$k=1$}
            \State {move distance $x_k$ away from $S$ (in any direction)}
        \Else
            \State {move distance $x_k$ away from $S$ in diretion of bomb, picking it up on the way}
        \EndIf
        \State {set down bomb}
        \State {return to source and $k \leftarrow k+1$}
        \EndFor
    \end{algorithmic}
\end{algorithm}
In the argument below we assume that given a strategy $X$ the adversary has the power to choose the speed of the fast robot and its initial distance from the source $S$. 

To ensure progress in searching, each trip away from the source should explore farther towards the perimeter than in the previous trip: this is formalized by the requirement that $x_k < x_{k+1}$ for all $k \geq 1$. Moreover, $\lim_{k\to \infty} x_k = +\infty$ (if not, the strategy could not solve all instances of the problem).

Consider a strategy $X$. Let the perimeter be at an unknown distance $D$. In each round $k$ for which the perimeter is not found the robot must return to the source and will therefore cover a length $2x_k$. The total length covered up to and including round $k$ will be equal to $2 \sum_{i=1}^k x_i$. If the perimeter is found during the next round the total distance covered by the robot will be $D+2 \sum_{i=1}^k x_i$.  The resulting competitive ratio will be equal to 
$$
\frac{D+2 \sum_{i=1}^k x_i}{D} = 1 + \frac{2 \sum_{i=1}^k  x_i}{D}
$$
Since the perimeter can be placed arbitrarily close to $x_k$ by an adversary it follows that the highest lower bound on the competitive ratio for this step will be equal to
$$
1 + \sup_{D> x_k} \frac{2 \sum_{i=1}^k  x_i}{D} = 
1 + \frac{2 \sum_{i=1}^k  x_i}{x_k} =
3 +  \frac{2\sum_{i=1}^{k-1} x_i}{x_k}
$$
It follows from the previous discussion that the resulting competitive ratio of the strategy $X$ will satisfy
\begin{equation}
\label{eq:CR}
CR_X = 3 +  \frac{2 \sup_{k \geq 1} \sum_{i=1}^{k-1} x_i}{x_k} .
\end{equation}

Observe that the lower bound obtained in Equation~\eqref{eq:CR} is valid for two robots provided the adversary can force the slow robot to execute the zigzag strategy. So we consider an optimal strategy $X=\{ x_1, x_2, \ldots, x_k, \ldots \}$. Let $x_k$ be the last move of this strategy with which we reach the destination perimeter. 

Now we are ready to complete our analysis in the {\tt InvisibleBoundary} model by proving the following:
\begin{theoremrep}\label{thm0-lb}
    The competitive ratio of every strategy $X$ solving the bomb squad problem in the {\tt InvisibleBoundary} model must satisfy $CR_X \geq 2 + \sqrt{5}$.  
\end{theoremrep}
\begin{proof}
    Consider the sum $s := \sum_{i=1}^{k-1} x_i$ which arises from the right-hand side of Equation~\eqref{eq:CR}. 
    Let $0 < \alpha < 1$ be a real number which is to be determined below.
    We distinguish two cases depending on the size of the sum $s$.

    \noindent{\bf Case 1:} $s\geq \alpha x_k$.\\ 
    From Inequality~\eqref{eq:CR} above we have that
    \begin{equation}
    \label{eq:CR1}
    CR \geq 3+ \frac{2s}{x_k} \geq 3 + 2 \alpha .
    \end{equation}

    \noindent{\bf Case 2:} $s < \alpha x_k$.\\
    The adversary places the robot $r_2$ at a distance $d_2$ so that $\frac {d_2}{v_2} \approx  s + \epsilon$ (so that the robots don't meet before $r_1$ leaves $S$ for the k-th time) and ensures the speed is such that $r_2$ is the robot that delivers optimally. 
    It is clear that the optimal algorithm is for robot $r_1$ to meet robot $r_2$ and hand the bomb which in turn delivers it to the perimeter. 
    The time it takes to do that is equal to 
    $$
    \frac{d_2}{1+v_2} + \frac{x_k - \frac{d_2}{1+v_2} }{v_2} = 
    \frac{d_2}{1+v_2} \left( 1 - \frac 1{v_2} \right) + \frac{x_k}{v_2} 
    \approx \frac{d_2}{v_2} + \frac{x_k}{v_2} = s + \epsilon + \frac{x_k}{v_2}, 
    $$
    where the last approximation follows from the fact that the adversary can make $d_2 , v_2$ arbitrarily large and at the same time maintain the constant ratio $\frac {d_2}{v_2} \approx  s + \epsilon$.
    However, the cost of the strategy $X$ will be at least $s + 2x_k$ (since the slow robot has the bomb but just missed the perimeter---placed at distance $D>x_k$---and has to return to the source) plus $\frac{x_k}{v_2}$ for the faster robot to deliver the bomb. It follows from the previous discussion that
    \begin{align}
    CR_X &\geq 
    \frac{s + \frac {x_k}{v_2} + 2x_k}{s + \epsilon + \frac {x_k}{v_2}}
    \approx  
    1 + \frac{2x_k}{s + \frac {x_k}{v_2}} 
    \geq \label{eq:CR2} 
    1 + \frac 2\alpha .
    \end{align}

    Using Inequalities~\eqref{eq:CR1}~and~\eqref{eq:CR2}, we obtain the lower bound $CR_X \geq \max \{ 3 + 2 \alpha, 1 + \frac 2\alpha  \}$. 
    The lower bound is maximized when $1 + \frac 2\alpha = 3 + 2 \alpha$ since $1 + \frac{2}{\alpha}$ is decreasing and $3+2\alpha$ is increasing on $0 \leq \alpha \leq 1$.
    If we multiply both sides of this equation by $\frac \alpha2$, we derive the quadratic $\alpha^2 + \alpha  -  1 = 0$ from which we obtain the solution $\alpha = \frac {-1+\sqrt{5}}{2}$. 
    The resulting lower bound on the competitive ratio is $CR_X \geq 3 + 2 \alpha = 2 + \sqrt{5}$.
    ~\qed
\end{proof}

\section{Conclusion}
\label{sec:Conclusion}

The main focus of the paper was to investigate algorithms for delivering a bomb to a safe location and compare the performance of online algorithms under several models which describe the knowledge the two robots have about each other and the environment (in this case the boundary). There are many interesting and challenging open problems remaining. For the two-robot case studied in the present paper, one can see the gaps remaining by glancing at the results displayed in Table~\ref{default}. An interesting class of problems arises in the multi-robot (more than two robots) case, where, generally, it is much harder to give tight performance bounds. Finally, it would be interesting to investigate algorithms that are resilient to faults that arise either from robot miscommunication or faults caused by the planar environment on which the robots operate (e.g. based on visibility obstructions, distance constraints, etc.) and/or are sensitive to energy consumption limitations.

\newpage

\bibliographystyle{splncs04}
\bibliography{refs}

\end{document}